# A Representation of Quantum Measurement in Nonassociative Algebras

Gerd Niestegge
Zillertalstrasse 39, 81373 Muenchen, Germany
gerd.niestegge@web.de

*Abstract.* Starting from an abstract setting for the Lüders-von Neumann quantum measurement process and its interpretation as a probability conditionalization rule in a non-Boolean event structure, the author derived a certain generalization of operator algebras in a preceding paper. This is an order-unit space with some specific properties. It becomes a Jordan operator algebra under a certain set of additional conditions, but does not own a multiplication operation in the most general case. A major objective of the present paper is the search for such examples of the structure mentioned above that do not stem from Jordan operator algebras; first natural candidates are matrix algebras over the octonions and other nonassociative rings. Therefore, the case when a nonassociative commutative multiplication exists is studied without assuming that it satisfies the Jordan condition. The characteristics of the resulting algebra are analyzed. This includes the uniqueness of the spectral resolution as well as a criterion for its existence, subalgebras that are Jordan algebras, associative subalgebras, and more different levels of compatibility than occurring in standard quantum mechanics. However, the paper cannot provide the desired example, but contribute to the search by the identification of some typical differences between the potential examples and the Jordan operator algebras and by negative results concerning some first natural candidates. The possibility that no such example exists cannot be ruled out. However, this would result in an unexpected new characterization of Jordan operator algebras, which would have a significant impact on quantum axiomatics since some customary axioms (e.g., power-associativity or the sum postulate for observables) might turn out to be redundant then.

## 1. Introduction

Starting from an abstract setting for the Lüders-von Neumann quantum measurement process and its interpretation as a probability conditionalization rule in a non-Boolean event structure [5], the author derived a certain generalization of operator algebras in a preceding paper [6], where two extreme cases were considered - the most general one without any multiplication operation and a set of additional conditions resulting in a Jordan operator algebra. The present paper studies a case between these two extremes; this is the case when a nonassociative commutative multiplication exists without assuming that it satisfies the Jordan condition $x^2 \circ (x \circ y) = x \circ (x^2 \circ y)$. This case is ideally suited for studying the question whether matrix algebras over the octonions or other nonassociative rings can provide examples of the structure mentioned above.

The characteristics of the resulting nonassociative algebra are analyzed. Although the complete algebra need not any more satisfy the Jordan condition, some subalgebras still do. If a spectral resolution exists, it is unique, and it exists for those elements which generate an associative subalgebra with positive squares. Moreover, different levels of compatibility are investigated, and there seem to be more different levels than occurring in standard quantum mechanics.

A major motivation of the present paper is to support the search for such examples of the structure mentioned above that do not stem from Jordan operator algebras. However, such examples have not been found yet; the paper contributes to the search by the identification of some typical differences between the potential examples and the Jordan operator algebras and by negative results concerning some first natural candidates.

The monograph [2] is recommended as reference for the theory of Jordan operator algebras; it also includes some basic material on order-unit spaces. A brief sketch of Jordan operator algebras and order unit spaces as far as needed in the present paper is provided in section 2 together with a theorem by Iochum and Loupias [3] that will be used in section 3. The structure of the nonassociative algebras under consideration is presented and studied in section 3. A certain type of "small" algebras is analyzed in section 4; they turn out to be identical with the spin factors or type $I_2$ factors from the theory of Jordan operator algebras. The different levels of compatibility are considered in section 5 and the matrix algebras in section 6.

## 2. Power-associative algebras

An algebra is called power-associative if each element $x$ lies in an associative subalgebra; this is equivalent to $x^n \circ x^m = x^{n+m}$ for $n, m \in \mathbb{N}$, where $x^n$ is inductively defined via $x^{n+1} = x \circ x^n$. Jordan algebras are always power-associative, but a power-associative algebra need not be a Jordan algebra. Jordan, von Neumann and Wigner [4] showed that the finite-dimensional formally real power-associative commutative algebras are Jordan algebras. Iochum and Loupias [3] extended this result to the infinite-dimensional case making use of the theory of JB and JBW algebras and order-unit spaces [2].

A JB algebra is a complete normed real Jordan algebra $M$ satisfying $\|a \circ b\| \leq \|a\| \, \|b\|$, $\|a^2\| = \|a\|^2$ and $\|a^2\| \leq \|a^2 + b^2\|$ for $a, b \in M$. A partial order relation $\leq$ on $M$ can then be derived by defining its positive cone as $\{a^2 : a \in M\}$. If $M$ is unital, we denote the identity by $\mathbb{I}$. A JB Algebra $M$ that owns a predual $M_*$ (i.e., $M$ is the dual space of $M_*$) is called a JBW algebra and is always unital. A JBW algebra can also be characterized as a JB algebra where each bounded monotone increasing net has a supremum in $M$ and a normal positive linear functional not vanishing in $a$ exists for each $a \neq 0$ in $M$ (i.e., the normal positive linear functionals are separating). A map is normal if it commutes with the supremum. It then turns out that the normal functionals coincide with the predual. The self-adjoint part of any W*-algebra (von Neumann algebra) equipped with the Jordan product $a \circ b := (ab + ba)/2$ is a JBW algebra.

An order-unit space is a partially ordered real vector space $L$ that contains an order-unit $\mathbb{I}$ and is Archimedean [2]. The order-unit $\mathbb{I}$ is positive and, for all $a \in L$, there is $t > 0$ such that $-t\mathbb{I} \leq a \leq t\mathbb{I}$. $L$ is Archimedean if $na \leq \mathbb{I}$ for all $n \in \mathbb{N}$ implies $a \leq 0$. An order-unit space $L$ has a norm given by $\|a\| = \inf\{t > 0 : -t\mathbb{I} \leq a \leq t\mathbb{I}\}$. Each $x \in L$ can be written as $x = a - b$ with positive $a, b \in A$ (e.g., choose $a = \|x\|\mathbb{I}$ and $b = \|x\|\mathbb{I} - x$). A positive linear functional $\rho : L \to \mathbb{R}$ on an order-unit space $L$ is norm continuous with $\|\rho\| = \rho(\mathbb{I})$ and, vice versa, a norm continuous linear functional $\rho$ with $\|\rho\| = \rho(\mathbb{I})$ is positive. Note that unital JB algebras are order unit spaces.

The order-unit space $L$ considered in the following is the dual space of a Banach space $V$ such that the unit ball of $L$ is compact in the weak-*-topology $\sigma(L, V)$. We will identify $\rho \in V$ with its canonical embedding in $V^{**} = L^*$. Then $L$ is monotone complete and $\rho(\sup x_\alpha) =$

lim$\rho(x_\alpha)$ holds for $\rho \in V$ and any bounded monotone increasing net $x_\alpha$ in $L$; in the operator algebra setting one would say that $\rho \in V$ is normal.

Iochum and Loupias [3] showed that, in the definition of a JB algebra, the Jordan condition can be replaced by power-associativity. This result will be used in section 3, and a slightly different proof is presented here because not only the result itself, but also a major part of this proof will be needed in section 3.

**Theorem 2.1:** (*Iochum/Loupias 1985*) *Suppose that $M$ is a power-associative commutative normed algebra over the real numbers with unit element $\mathbb{1}$ such that $\|x \circ y\| \leq \|x\| \|y\|$, $\|x^2\| = \|x\|^2$ and $\|x^2\| \leq \|x^2+y^2\|$ for $x,y \in M$. Then $M$ is a Jordan algebra.*

*Proof*: Without loss of generality we may assume that $M$ is norm complete. Define $M_+:=\{y^2: y \in M\}$ and denote by $C(x)$ the norm closed subalgebra generated by $x \in M$ and $\mathbb{1}$. Then $C(x)$ is an associative JB algebra and thus isomorphic to the algebra of continuous functions on some compact Hausdorff space. Therefore we get for $x \in M$ with $\|x\| \leq 1$ that $x \in M_+$ if and only if $\|\mathbb{1}-x\| \leq 1$. This implies that $M_+$ is a convex cone. Since $x^2=-y^2$ with $x,y \in M$ implies $0=\|x^2+y^2\|$ and thus $0=\|x^2\|=\|x\|^2$ such that $x=0$, $M_+$ defines a partial ordering on $M$ making $M$ an order-unit space with order unit $\mathbb{1}$.

Thus a linear functional $\rho$ in the dual space $M^*$ with $\|\rho\|=\rho(\mathbb{1})$ is positive. Then we have $\mu(x^2) \geq 0$ for $x \in M$ and $\mu \in S := \{\rho \in M^*: \|\rho\|=\rho(\mathbb{1})=1\}$. The Cauchy-Schwarz inequality yields $(\mu(x \circ y))^2 \leq \mu(x^2)\mu(y^2)$ for $x,y \in M$, $\mu \in S$, and $(\mu(x))^2 \leq \mu(x^2)$ with $y=\mathbb{1}$. Moreover, each $\rho \in M^*$ has the shape $\rho=s\mu_1-t\mu_2$ with $\mu_1,\mu_2 \in S$ and $s,t \geq 0$. Now consider the seminorms $x \to \mu(x^2)^{1/2}$, $\mu \in S$, on $M$ and the topology defined by them on $M$ which is called the *s*-topology. Norm convergence implies *s*-convergence. Due to the Cauchy-Schwarz inequality, *s*-convergence implies $\sigma(M,M^*)$-convergence and the product $x \circ y$ is *s*-continuous separately in each factor. We shall now prove that the product $x \circ y$ is jointly *s*-continuous in both factors on bounded subsets of $M$.

Let $x_\alpha$ and $y_\beta$ be two bounded nets in $M$. First suppose that the net $x_\alpha$ *s*-converges to 0. Considering $C(x_\alpha)$ we find that $0 \leq (x_\alpha^2)^2 \leq \|x_\alpha^2\| x_\alpha^2$ holds for each $\alpha$; therefore the net $x_\alpha^2$ *s*-converges to 0. If furthermore the net $y_\beta$ *s*-converges to 0, the identity $x \circ y = ((x+y)^2 - x^2 - y^2)/2$ implies that the net $x_\alpha \circ y_\beta$ *s*-converges to 0. Now suppose that the two nets $x_\alpha$ and $y_\beta$ *s*-converge to $x_o$ and $y_o$, respectively. Then use the identity $x_o \circ y_o - x_\alpha \circ y_\beta = (x_o-x_\alpha) \circ y_o + (x_\alpha-x_o) \circ (y_o-y_\beta) + x_o \circ (y_o-y_\beta)$ to conclude that $x_\alpha \circ y_\beta$ *s*-converges to $x_o \circ y_o$.

Furthermore, consider the second dual $M^{**}$ and assume that $M$ is canonically embedded in $M^{**}$. Let $N$ comprise all those elements of $M^{**}$ that are the $\sigma(M,M^*)$-limit of a bounded net in $M$ which is a Cauchy net with respect to the seminorms defining the *s*-topology. Then the product $\circ$ has an *s*-continuous extension to $N$ and $N$ is power-associative. Moreover, $\mu(x^2) \geq 0$ for $\mu \in S$, $x \in N$. Therefore, the Cauchy-Schwarz inequality again holds and, since $\|x\|=\sup\{|\mu(x)|: \mu \in S\}$ for $x \in N$, $N$ inherits from $M$ the properties $\|x^2\| = \|x\|^2$ and $\|x^2+y^2\| \geq \|x^2\|$ for $x,y \in N$. Note that $M^{**}$ does not automatically inherit these properties since the map $x \to x^2$ is not $\sigma(M,M^*)$-continuous. An element $x \in N$ is positive iff $\mu(x) \geq 0$ for $\mu \in S$, or iff $x=a^2$ for some $a \in N$.

Again, the norm closed subalgebra $C(a)$ generated by some $a \in N$ and $\mathbb{1}$ is an associative JB algebra; therefore $a^2 \leq \|a\|a$ holds for positive $a \in N$. If now $x_\alpha$ is a bounded monotone increasing net in $N$, then $x_\alpha$ $\sigma(M^{**},M^*)$-converges to sup $x_\alpha$ in $M^{**}$. Since $(x-y)^2 \leq \|x-y\|(x-y)$ for $y \leq x$ in $N$, the net $x_\alpha$ *s*-converges to sup $x_\alpha$ such that sup $x_\alpha \in N$. Therefore $N$ is monotone complete and the restrictions of the positive elements of $V$ provide a separating family of normal functionals.

In the same way, one can conclude that the *s*-closed subalgebra $W(x) \subseteq N$ generated by some $x \in N$ and $\mathbb{1}$ is monotone complete with a separating family of normal functionals such that it becomes an associative JBW algebra. Then the spectral theorem holds and $x$ can be norm-approximated by elements having the shape $a = \Sigma t_k e_k$ with real numbers $t_k$, idempotent elements $e_1,...,e_n$ in $W(x)$ and $e_k \circ e_l = 0$ for $k \neq l$. By a result in [7], the identity $e \circ (f \circ y) = f \circ (e \circ y)$ holds in any power-associative algebra for idempotent elements $e$ and $f$ with $e \circ f = 0$ and any $y$. Therefore

$$a^2 \circ (a \circ y) = \Sigma_k \Sigma_l t_k^2 t_l e_k \circ (e_l \circ y) = \Sigma_k \Sigma_l t_k^2 t_l e_l \circ (e_k \circ y) = a \circ (a^2 \circ y)$$

and thus $x^2 \circ (x \circ y) = x \circ (x^2 \circ y)$ for $x,y \in N$ and particularly for $x,y$ in the subalgebra $M$.        q.e.d.

## 3. Nonassociative algebras

The structure that we will study is motivated by the results in [6], where an order-unit space with a specific type of positive projections was derived from an abstract setting for conditional probabilities and the Lüders-von Neumann quantum measurement. A projection is a linear map $U: A \to A$ on the order-unit space $A$ with $U^2 = U$. Examples of this structure are the JBW algebras, the finite-dimensional version of which are the formally really Jordan algebras. In these cases the specific positive projections have the shape $U_e x = \{e,x,e\}$ with an idempotent element $e$, where $\{a,b,c\} := a \circ (b \circ c) - b \circ (c \circ a) + c \circ (a \circ b)$ denotes the so-called triple product. If $e$ is idempotent, $\{e,x,e\}$ becomes $2e \circ (e \circ x) - e \circ x$. Therefore, in the JBW case, there is a close relation between the specific positive projections and the Jordan product. The idea behind the following assumptions is to keep the connection of the positive projections with a nonassociative product without imposing any further restrictions; particularly the product need not satisfy the Jordan condition.

For any set $K$ in an order-unit space $A$ with predual $V$ denote by $\overline{\mathrm{lin}}\, K$ the $\sigma(A,V)$-closed linear hull of $K$.

**Assumptions 3.1:** (i) *A is an order-unit space.*
(ii) *A is the dual of the Banach space V.*
(iii) *A is a real algebra with the (not necessarily associative) commutative multiplication* $\circ$.
(iv) *The element* $\mathbb{1}$ *in A is the order unit and the identity for the multiplication.*
(v) $||x \circ y|| \leq ||x||\, ||y||$ *for* $x,y \in A$.
(vi) *The product* $x \circ y$ *is* $\sigma(A,V)$-*continuous in x with y fixed as well as in y with x fixed.*
*We define* $E := \{e \in A: e \circ e = e\}$, $U_e x := \{e,x,e\}$ *for* $e \in E$, $x \in A$, *and* $S := \{\mu \in V: ||\mu|| = \mu(\mathbb{1}) = 1\}$.
(vii) *The linear map* $U_e: A \to A$ *is a positive projection with* $U_e A = \overline{\mathrm{lin}}\, \{f \in E: f \leq e\}$ *for each* $e \in E$.
(viii) *If* $\mu \in S$ *and* $e \in E$ *with* $\mu(e) = 1$, *then* $\mu$ *is invariant under* $U_e$ (*i.e.,* $\mu = \mu U_e$).

In the remaining part of the present paper, $A$ shall always satisfy all the conditions (i) - (viii). From $0 \leq U_e \mathbb{1} = e$ for $e \in E$ we get that the elements of $E$ are positive. Since $\mathbb{1} - e \in E$ is also positive, we have $0 \leq e \leq \mathbb{1}$. With the orthocomplementation $e' := \mathbb{1} - e$, the set $E$ becomes an orthomodular partially ordered set. Two elements $e,f \in E$ are called orthogonal if $f \leq e'$; then $U_e f = 0$ since $0 \leq U_e f \leq U_e e' = \{e, \mathbb{1} - e, e\} = 0$. Moreover, $U_e U_{e'} = U_{e'} U_e = 0$.

As already in Refs. [5] and [6], we interpret the set $E$ consisting of the idempotent elements of $A$ as a generalized non-Boolean event structure and call the elements of $E$ events. This is the viewpoint of probability theory. From another viewpoint, $E$ could also be called a quantum logic and the elements of $E$ could be called propositions.

Some more background information and motivation for the above conditions (i), (ii), (vii) and (viii) can be found in [6] where they were derived from a few very basic assumptions concerning events, states and particularly the conditional probabilities. The new conditions (iii), (iv), (v) and (vi) mean that the same relation to a nonassociative product ∘ is assumed as we find it in the JBW algebras: The events become idempotent elements with regard to this product and the positive projections have the shape $U_e x = \{e,x,e\} = 2e \circ (e \circ x) - e \circ x$. However, we do neither assume here that the product satisfies the Jordan condition or that it is power-associative nor that the conditions $\|a^2\|=\|a\|^2$ or $\|a^2\|\leq\|a^2+b^2\|$ hold for the norm.

Note that the link to quantum measurement and conditional probabilities is the formula $\mu(f|e) = \mu(U_e f)/\mu(e)$ for $\mu \in S$, $e,f \in E$ with $\mu(e)>0$ (see [6]). Here $\mu(f|e)$ denotes the conditional probability of the event $f$ under another event $e$ in the state $\mu$. In the quantum measurement setting, $\mu(f|e)$ is the probability that a second measurement provides the result $f$ after a first measurement has already been performed and has provided the result $e$, assuming that the physical system under consideration is in the state $\mu$. In a special Jordan algebra (e.g., the self-adjoint part of a W*-algebra), $U_e f = \{e,f,e\}$ becomes $efe$, which reveals the connection to the Lüders - von Neumann measurement process in the customary Hilbert space model of quantum mechanics.

The structure of the algebra $A$ is designed in such a way that it owns all those properties of a JBW algebra that are necessary to make the map $f \to \mu(\{e,f,e\})/\mu(e)$ a unique conditional probability within the class $S$ of normalized positive linear functionals on $A$. The situation in [5,6] was a little different. There unique conditional probabilities within the normalized positive additive functions on $E$ were considered, which requires a Gleason type theorem making sure that these functions on $E$ have linear extensions to $A$.

We shall now study subalgebras of $A$ and identify conditions that make them Jordan algebras. Since the intersection of any family of monotone closed subalgebras is a monotone closed subalgebra, there is a smallest monotone closed subalgebra containing any given subset of $A$; it is called the monotone closed subalgebra generated by the subset. Note that the following theorem does not require the conditions (vii) and (viii) of the assumptions 3.1.

**Theorem 3.2:** *Suppose that $M$ is a power-associative subalgebra of $A$ with $\mathbb{I} \in M$ and $x^2 \geq 0$ for each $x \in M$. Then $M$ is a Jordan algebra, its norm closure is a JB algebra, and the monotone closed subalgebra that $M$ generates is a JBW algebra.*

*Proof.* Since $\mu(x^2) \geq 0$ for $x \in M$ and $\mu \in S$, the Cauchy-Schwarz inequality yields $(\mu(x \circ y))^2 \leq \mu(x^2)\mu(y^2)$ for $x,y \in M$ and $(\mu(x))^2 \leq \mu(x^2)$ with $y=\mathbb{I}$. Therefore $\|x\|^2 \leq \|x\|^2 = \sup\{(\mu(x))^2: \mu \in S\}$ $\leq \sup\{\mu(x^2): \mu \in S\} = \|x^2\|$ such that $\|x^2\| = \|x\|^2$ for $x \in M$. Moreover $\|x^2+y^2\| = \sup\{\mu(x^2)+\mu(y^2): \mu \in S\} \geq \sup\{\mu(x^2): \mu \in S\} = \|x^2\|$ for $x,y \in M$. By Theorem 2.1, $M$ is a Jordan algebra and its norm closure a JB algebra.

As in the proof of Theorem 2.1 consider the *s*-topology again and let $N$ comprise all those elements of $A$ that are the $\sigma(A,V)$-limit of a bounded net in $M$ which is a Cauchy net with respect to the seminorms defining the *s*-topology. As with Theorem 2.1 now conclude that $N$ is a JBW algebra. The monotone closed subalgebra generated by $M$ is contained in $N$ and thus a JBW algebra as well.                                                                                          q.e.d.

**Lemma 3.3:** (i) *Two elements $e$ and $f$ in $E$ are orthogonal (i.e., $f \leq e' = \mathbb{I} - e$) iff $U_e f = f$, or iff $e \circ f = 0$. They satisfy $e \leq f$ iff $e \circ f = e$.*
(ii) *If $e_n$ is an orthogonal sequence in $E$, then $\Sigma e_n \in E$.*

*Proof.* (i) If $f \leq e'$, then $U_e f = f$ holds since $f \in U_e A$. If $U_e f = f$, then $U_e f = 0$ and the identity $e \circ f = (f + U_e f - U_e f)/2$ implies $e \circ f = 0$. If $e \circ f = 0$, then $\mathbb{I} - e - f = (\mathbb{I} - e - f)^2$ such that $\mathbb{I} - e - f \in E$ and thus $\mathbb{I} - e - f \geq 0$ or $f \leq e'$. Moreover, we have $e \leq f$ iff $e$ and $f'$ are orthogonal, and this is equivalent to $0 = e \circ f' = e \circ (\mathbb{I} - f) = e - e \circ f$.

(ii) The sum $\Sigma e_n$ exists in $A$ due to the monotone completeness of $A$ and converges with regard to the $\sigma(A,V)$-topology. By (i) the orthogonality implies $e_n \circ e_m = 0$ for $n \neq m$. Thus $(\Sigma e_n) \circ e_m = e_m$ for each $m$ and $(\Sigma e_n) \circ (\Sigma e_n) = \Sigma e_n$.      q.e.d.

It follows from the above lemma that the algebra $A$ is associative if and only if $E$ is a Boolean lattice (or Boolean algebra). If $A$ is associative, $e \circ f \in E$ for $e, f \in E$ and $E$ becomes a Boolean lattice with $e \wedge f = e \circ f$. If $E$ is a Boolean lattice, then any two elements $e$ and $f$ in $E$ can be decomposed as $e = d_1 + d_2$ and $f = d_2 + d_3$ with orthogonal elements $d_1, d_2, d_3 \in E$. Then $e \circ f = d_2 = e \wedge f$. Therefore $d \circ (e \circ f) = d \wedge e \wedge f = (d \circ e) \circ f$ for any $d, e, f \in E$ and $A$ becomes associative since it is generated by $E$.

A spectral measure $X$ allocates to each Borel measurable subset $B$ of the real numbers $\mathbb{R}$ an idempotent element $e_B$ in $E$ such that the map $B \to e_B$ is $\sigma$-additive and $e_B = \mathbb{I}$ for $B = \mathbb{R}$. If $\mu \in S$, then $B \to \mu(e_B)$ becomes a probability measure over $\mathbb{R}$ which is denoted by $\mu^X$. The spectral measure $X$ is called a spectral resolution of $x \in A$ if the measure integral $\int t \, d\mu^X$ coincides with $\mu(x)$ for all $\mu \in S$. Such an $x$ exists in $A$ and is uniquely determined for each bounded spectral measure $X$ [6]. However, not each $x$ in $A$ has a spectral resolution.

It follows from Theorem 3.2 that elements of $A$ that lie in a power-associative subalgebra with positive squares have a spectral resolution and that the spectral resolution is uniquely determined in the generated JBW subalgebra. We shall now see that it is uniquely determined in $A$.

**Proposition 3.4:** *Suppose that an element $x \in A$ has a spectral resolution. Then its spectral measure $X$ is uniquely determined in A. Moreover, the norm closed subalgebra generated by $x$ and $\mathbb{I}$ is an associative JB algebra and the monotone closed subalgebra generated by $x$ and $\mathbb{I}$ is an associative JBW algebra.*

*Proof.* Suppose that $x \in A$ has a spectral resolution. The spectral measure must then be bounded and $x$ can uniformly be approximated by elements having the shape $a = \Sigma t_k e_k$ with real numbers $t_k$, idempotent elements $e_1, ..., e_n$ in $A$ and $e_k \circ e_l = 0$ for $k \neq l$. Since elements with this shape are power-associative with $a^m = \Sigma t_k^m e_k$, we get that $x$ is power-associative and that $\int t^n d\mu^X = \mu(x^n)$ for all $\mu \in S$. Because the moments of a probability distribution $\mu^X$ uniquely determine the distribution (this follows from the Fourier transformation), we get that $\mu(e_B)$ is uniquely determined for all $\mu \in S$ and thus $e_B$ is uniquely determined in $A$ for every Borel set $B$.

Moreover, $\mu((p(x))^2) = \int (p(t))^2 d\mu^X \geq 0$ for $\mu \in S$ and any polynomial $p$ such that $(p(x))^2 \geq 0$. The subalgebra generated by $x$ and $\mathbb{I}$ is associative and the squares of its elements are positive. As in the proof of Theorem 3.2 now conclude that its norm closure is an associative JB algebra and that the generated monotone closed subalgebra is an associative JBW algebra.      q.e.d.

Real-valued observables can be defined as spectral measures and those elements in $A$ which own a spectral resolution can be identified with real-valued observables therefore [5,6]. We shall now see that $A$ becomes a JBW algebra if each element in $A$ represents an observable.

**Corollary 3.5:** *If each element in A has a spectral resolution, then A is a JBW algebra.*

*Proof.* By Proposition 3.4 *A* is power-associative and the squares of its elements are positive. Then apply Theorem 3.2.                                                                 q.e.d.

With Theorem 3.2 and Proposition 3.4, an element *x* in *A* has a spectral resolution within the idempotent elements of *A* if and only if *x* lies in an associative subalgebra such that $y^2 \geq 0$ holds for all *y* in this subalgebra. Associative JB or JBW algebras are also called abelian and are identical with the self-adjoint parts of the abelian C*-algebra and abelian W*-algebras, respectively. The abelian subalgebras play an important role in the theory of operator algebras.

In the algebra *A*, however, one can study two further potential properties of its subalgebras - power-associativity of a subalgebra and the positivity of the square of each element in a subalgebra. The subalgebras with both these properties become Jordan algebras, while *A* itself need not be a Jordan algebra. Concerning the search for an example that satisfies the assumptions 3.1, but is not a JB algebra, this means that such an example would have to violate at least one of the two properties: power-associativity or positivity of the squares.

**4. "Small" algebras**

An event $0 \neq e \in E$ is called minimal, if there is no other nonzero event *f* with $f \leq e$; then $U_e A = \mathbb{R} e$. In this section, we are going to study algebras where all nontrivial events are minimal. Such algebras can be considered small in the sense that any orthogonal family of nonzero events cannot contain more than two elements. They represent the most simple case which is possible with the assumptions 3.1.

**Theorem 4.1:** *Suppose that A and $E=\{e \in A: e^2=e\}$ satisfy the assumptions* 3.1 *and that each element in E which differs from* 0 *and* $\mathbb{I}$ *is minimal. Then A is a JB algebra.*

*Proof.* Suppose that *e* and *f* are any minimal events different from $\mathbb{I}$. Then *e'* and *f'* are minimal as well. Furthermore $U_e f = \lambda e$ and $U_{e'} f = \lambda' e'$ with some $\lambda, \lambda' \in [0,1]$. In a first step, we show that $\lambda + \lambda' = 1$.

For any $r,s,t \in \mathbb{R}$, consider the linear combination $x := re + se' + tf$ in *A*. From $2e \circ f = f + U_e f - U_{e'} f = f + \lambda e - \lambda' e'$ and $2e' \circ f = f - \lambda e + \lambda' e'$, we get $2f \circ x = (r-s)(\lambda e - \lambda' e') + (2r+2s+t)f$, $2f \circ (f \circ x) = ((r-s)/2)(\lambda(f + \lambda e - \lambda' e') - \lambda'(f - \lambda e + \lambda' e')) + (2r+2s+t)f$ and $U_f x = \{f,x,f\} = 2f \circ (f \circ x) - f \circ x = ((r-s)/2)(\lambda + \lambda' - 1)(\lambda e - \lambda' e') + ((2r+2s+t)/2)f$. Since $U_f x \in \mathbb{R} f$ must hold for all *r,s,t*, either $\lambda + \lambda' = 1$ or $\lambda e = \lambda' e'$. The second case implies $\lambda = \lambda' = 0$ (multiply both sides first with *e* and then with *e'*) and $e \circ f = f/2$. Then $f' \circ e = (\mathbb{I} - f) \circ e = e - f \circ e = e - f/2$, $f' \circ (f' \circ e) = f' \circ e$ and $U_f e = 2f' \circ (f' \circ e) - f' \circ e = f' \circ e = e - f/2$ such that $e \in \mathbb{R} f \oplus \mathbb{R} f'$. Therefore $e = f$ and $\lambda = 1$, or $e = f'$ and $\lambda' = 1$, resulting in a contradiction to $\lambda = \lambda' = 0$. Therefore, only the case $\lambda + \lambda' = 1$ can occur.

Now consider any three minimal events *d,e,f* - each one different from $\mathbb{I}$ - and define $x := d - U_e d - U_{e'} d$, $y := f - U_e f - U_{e'} f$. In a second step, it is shown that $x \circ y \in \mathbb{R} \mathbb{I}$.

From the first step, we get $U_e f = \alpha e$, $U_{e'} f = (1-\alpha)e'$, $U_e d = \beta e$, $U_{e'} d = (1-\beta)e'$, $U_f d = \gamma f$ and $U_{f'} d = (1-\gamma)f'$ with some $\alpha, \beta, \gamma \in [0,1]$. Then $x = d - \beta e - (1-\beta)e' = d + (1-2\beta)e - (1-\beta)\mathbb{I}$ and $y = f - \alpha e - (1-\alpha)e' = f + (1-2\alpha)e - (1-\alpha)\mathbb{I}$. Therefore

$$x \circ y = d \circ f + (1-2\alpha)d \circ e - (1-\alpha)d + (1-2\beta)e \circ f + (\alpha+\beta-1)e - (1-\beta)f + (1-\alpha)(1-\beta)\mathbb{I}.$$

Using the identities $d \circ f = (d+U_f d - U_{f'}d)/2 = (d+\gamma f-(1-\gamma)f')/2 = (d+f-(1-\gamma)\mathbb{1})/2$, $d \circ e = (d+U_e d - U_{e'}d)/2 = (d+e-(1-\beta)\mathbb{1})/2$ and $e \circ f = (f+U_e f - U_{e'}f)/2 = (f+e-(1-\alpha)\mathbb{1})/2$ we get

$$\begin{aligned} x \circ y &= (d+f-(1-\gamma)\mathbb{1})/2 + (1-2\alpha)(d+e-(1-\beta)\mathbb{1})/2 - (1-\alpha)d + (1-2\beta)(f+e-(1-\alpha)\mathbb{1})/2 \\ &\quad + (\alpha+\beta-1)e - (1-\beta)f + (1-\alpha)(1-\beta)\mathbb{1} \\ &= d/2 + (1-2\alpha)d/2 - (1-\alpha)d + f/2 + (1-2\beta)f/2 - (1-\beta)f + (1-2\alpha)e/2 + (1-2\beta)e/2 \\ &\quad + (\alpha+\beta-1)e - (1-\gamma)\mathbb{1}/2 - (1-2\alpha)(1-\beta)\mathbb{1}/2 - (1-2\beta)(1-\alpha)\mathbb{1}/2 + (1-\alpha)(1-\beta)\mathbb{1} \\ &= (\alpha/2 + \beta/2 + \gamma/2 - \alpha\beta - 1/2)\mathbb{1} \in \mathbb{R}\mathbb{1}. \end{aligned}$$

In the third step, consider any two elements $a,b \in A$ and define $x := a - U_e a - U_{e'}a$, $y := b - U_e b - U_{e'}b$. Since $a$ and $b$ both are linear combinations of minimal events, we get from the second step that $x \circ y \in \mathbb{R}\mathbb{1}$. Therefore $x^2 = \lambda \mathbb{1}$ with $\lambda \in \mathbb{R}$. It is shown in the fourth step that then $\lambda \geq 0$ and $\lambda > 0$ for $x \neq 0$.

Suppose $\lambda < 0$. Without loss of generality assume $x^2 = -\mathbb{1}$ (if this is not the case, replace $x$ by $|\lambda|^{-1/2}x$) and define $f := \frac{1}{2}(1-5^{1/2})e + \frac{1}{2}(1+5^{1/2})e' + x$. From $e \circ x = \frac{1}{2}(x+U_e x - U_{e'}x)$ and $U_e x = U_{e'}x = 0$ we get $e \circ x = \frac{1}{2}x$ and $e' \circ x = (\mathbb{1}-e) \circ x = \frac{1}{2}x$. Therefore $f^2 = f$ and $U_e f = \frac{1}{2}(1-5^{1/2})e$. However, since $f \in E$, we have $U_e f \geq 0$, resulting in a contradiction to $\frac{1}{2}(1-5^{1/2}) < 0$.

Now suppose $\lambda = 0$. Then $x^2 = 0$, and $e \circ x = \frac{1}{2}x$ implies $(e+sx)^2 = e+sx$ such that $e+sx \in E$ and $e+sx \leq \mathbb{1}$ for all $s \in \mathbb{R}$. Therefore $sx \leq \mathbb{1}$ for all $s \in \mathbb{R}$ and $x = 0$.

Finally, we show that $A$ is power-associative and that the elements of $A$ have positive squares. An application of Theorem 3.2 then yields that $A$ is a JB algebra.

Since $A = U_e A \oplus U_{e'}A \oplus \{a - U_e a - U_{e'}a : a \in A\} = \mathbb{R}e \oplus \mathbb{R}e' \oplus \{a - U_e a - U_{e'}a : a \in A\}$, we have $A = \mathbb{R}\mathbb{1} \oplus H$ with $H := \mathbb{R}(e-e') \oplus \{a - U_e a - U_{e'}a : a \in A\}$. From $(e-e')^2 = \mathbb{1}$ and $(e-e') \circ x = \frac{1}{2}x - \frac{1}{2}x = 0$ for $x = a - U_e a - U_{e'}a$ with $a \in A$, it follows that $x \circ y \in \mathbb{R}\mathbb{1}$ for $x, y \in H$. Moreover, $x^2 = \lambda \mathbb{1}$ with $\lambda > 0$ for $x \neq 0$.

Suppose $a = s\mathbb{1} + x \in A$ with $s \in \mathbb{R}$ and $x \in H$. If $x = 0$, $a$ lies in the associative subalgebra $\mathbb{R}\mathbb{1}$ and $a^2 = s^2 \mathbb{1}$ is positive. Now consider the case $x \neq 0$. Since $x^2 = \lambda \mathbb{1}$ with $\lambda > 0$, we can define $d := \frac{1}{2}(\mathbb{1} + \lambda^{-1/2}x)$. Then $d^2 = d \in E$, $d' = \frac{1}{2}(\mathbb{1} - \lambda^{-1/2}x)$ and $a = (s+\lambda^{1/2})d + (s-\lambda^{1/2})d'$. Therefore $a$ lies in the associative subalgebra $\mathbb{R}d \oplus \mathbb{R}d'$ and $a^2 = (s+\lambda^{1/2})^2 d + (s-\lambda^{1/2})^2 d'$ is positive.      q.e.d.

In Theorem 4.1, $A$ is associative if $E = \{0, \mathbb{1}\}$ or $E = \{0, e, e', \mathbb{1}\}$ with some event $e$; in all other cases, $A$ is a so-called spin factor or type $I_2$ JBW factor (definitions can be found in [2]). Note that the dimension of a spin factor need not be finite; indeed there is a spin factor of dimension $n$ for each cardinal number $n \geq 3$, including the infinite cardinal numbers.

## 5. Compatibility

The following two lemmas concern orthogonal idempotent elements and will be needed for the investigation of the different notions of compatibility in this section and for the study of the matrix algebras in the next section.

**Lemma 5.1:** *Under the assumptions* 3.1, *suppose that $e$ and $f$ are two orthogonal elements of E. Then $U_{e'}U_{f'} = U_{(e+f)'} = U_{f'}U_{e'}$.*

*Proof.* Suppose $x \in [0, \mathbb{I}]$. Then $0 \leq U_{e'}U_{f'}x \leq U_{e'}U_{f'}\mathbb{I} = U_{e'}f' = e'\cdot f = (e+f)'$. Therefore $\mu U_{e'}U_{f'} = 0 = \mu U_{(e+f)'}$ for $\mu \in S$ with $\mu((e+f)')=0$. Now consider $\mu \in S$ with $\mu((e+f)') > 0$ and define $\nu := \mu U_{e'}U_{f'}/\mu((e+f)') \in S$. From $(e+f)' \leq e'$ and $(e+f)' \leq f'$, we get $U_{e'}U_{f'}(e+f)' = (e+f)'$ and $\nu((e+f)') = 1$ such that $\nu = \nu U_{(e+f)'}$. Since $U_{e'}U_{(e+f)'} = U_{(e+f)'} = U_{f'}U_{(e+f)'}$, we have $\nu = \mu U_{(e+f)'}/\mu((e+f)')$ and thus $\mu U_{e'}U_{f'} = \mu U_{(e+f)'}$. Therefore $U_{e'}U_{f'} = U_{(e+f)'}$. In the same way we get $U_{f'}U_{e'} = U_{(e+f)'}$.     q.e.d.

**Lemma 5.2:** *Under the assumptions* 3.1, *suppose that $e$ and $f$ are two orthogonal elements of E. Then $a \circ (b \circ x) = b \circ (a \circ x)$ for $a \in U_e A$, $b \in U_f A$ and $x \in A$.*

*Proof.* From the preceding lemma we have $U_{e'}U_{f'} = U_{f'}U_{e'}$. In [6] it was shown that $U_e U_f = U_f U_e = 0$, $U_{e'}U_f = U_f U_{e'} = U_f$ and $U_{f'}U_e = U_e U_{f'} = U_e$. Therefore $U_e$, $U_f$, $U_{e'}$, $U_{f'}$ commute pairwise. The identities $e \circ y = (y + U_e y - U_{e'} y)/2$ and $f \circ y = (y + U_f y - U_{f'} y)/2$ ($y \in A$) then imply $e \circ (f \circ x) = f \circ (e \circ x)$ for $x \in A$.

Since this holds for all orthogonal elements in $E$, we have $g \circ (h \circ x) = h \circ (g \circ x)$ for $g, h \in E$ with $g \leq e$ and $h \leq f$. Therefore $a \circ (b \circ x) = b \circ (a \circ x)$ for $a \in U_e A = \overline{\text{lin}}\{g \in E: g \leq e\}$ and $b \in U_f A = \overline{\text{lin}}\{h \in E: h \leq f\}$.     q.e.d.

Each condition in the following proposition represents a certain degree of compatibility; the first conditions represent a rather weak type of compatibility and the last ones a rather strong type. The proposition analyzes the precise logical relations among all the different conditions.

**Proposition 5.3:** *Under the assumptions* 3.1, *consider the following eleven conditions for a pair of events $e, f \in E$.*
(i) $f = U_e f + U_{e'} f$ (i.e., $\mu(f) = \mu(e)\mu(f|e) + \mu(e')\mu(f|e')$ for all $\mu \in S$).
(ii) $e \circ (e \circ f) = e \circ f$.
(iii) $f = U_e f + U_{e'} f$ and $e = U_f e + U_{f'} e$.
(iv) $e \circ (e \circ f) = f \circ (e \circ f) = e \circ f$.
(v) $U_e f = e \circ f = U_f e$.
(vi) $U_a b = U_b a$ for $a, b \in \{e, e', f, f'\}$ (i.e., $\mu(a)\mu(b|a) = \mu(b)\mu(a|b)$ for $a, b \in \{e, e', f, f'\}$ and $\mu \in S$).
(vii) $U_a U_b = U_b U_a$ for $a, b \in \{e, e', f, f'\}$.
(viii) $e \circ (f \circ x) = f \circ (e \circ x)$ for $x \in A$ (i.e., $e$ and $f$ operator-commute).
(ix) $e \circ f$, $e' \circ f$, $e \circ f'$ and $e' \circ f'$ lie in $E$.
(x) $e$ and $f$ lie in an associative subalgebra.
(xi) *There are three orthogonal elements $d_1, d_2, d_3 \in E$ such that $e = d_1 + d_2$ and $f = d_2 + d_3$.*
*Then the following logical relations hold among these conditions:* (i) $\Leftrightarrow$ (ii) $\Leftarrow$ (iii) $\Leftrightarrow$ (iv) $\Leftrightarrow$ (v) $\Leftrightarrow$ (vi) $\Leftarrow$ (vii) $\Leftrightarrow$ (viii) $\Leftarrow$ (ix) $\Leftrightarrow$ (x) $\Leftrightarrow$ (xi).

*Proof.* From the identity $U_e f + U_{e'} f = \{e,f,e\} + \{\mathbb{I}-e, f, \mathbb{I}-e\} = 2\{e,f,e\} + f - 2e \circ f = 4e \circ (e \circ f) - 4e \circ f + f$ we immediately get the equivalence of (i) and (ii). The implication (i) $\Leftarrow$ (iii) is obvious. The equivalence of (iii) and (iv) follows from the one for (i) and (ii), the one of (iv) and (v) from the identities $U_e f = \{e,f,e\} = 2e \circ (e \circ f) - e \circ f$ and $U_f e = 2f \circ (e \circ f) - e \circ f$.

Now suppose (v); then we also have (iv) and $e' \circ (e' \circ f) = f - 2e \circ f + e \circ (e \circ f) = f - e \circ f = e' \circ f$ such that $U_{e'}f = 2e' \circ (e' \circ f) - e' \circ f = e' \circ f$ and $U_f e' = U_f(\mathbb{I}-e) = f - U_f e = f - e \circ f = e' \circ f = U_{e'}f$. In the same way conclude that $U_{f'}e = e \circ f' = U_e f'$. Furthermore $U_{e'}f' = e' - U_e f = e' - e' \circ f = e' \circ f'$ and $U_{f'}e' = f' - U_{f'}e = f' - e \circ f' = e' \circ f' = U_{e'}f'$. The remaining cases for (vi) are trivial. Condition (vi) implies (iii) via $U_e f + U_{e'} f = U_f e + U_f e' = U_f(e+e') = U_f \mathbb{1} = f$ and $U_f e + U_{f'} e = e$ in the same way, and we have the equivalence of (iii), (iv), (v) and (vi).

Condition (vii) implies (vi) via $U_a b = U_a U_b \mathbb{1} = U_b U_a \mathbb{1} = U_b a$, and (viii) follows from (vii) by the identities $e \circ y = (y + U_e y - U_{e'} y)/2$ and $f \circ y = (y + U_f y - U_{f'} y)/2$ ($y \in A$). Vice versa, (viii) implies $a \circ (b \circ x) = b \circ (a \circ x)$ for $x \in A$ and $a,b \in \{e,e',f,f'\}$; this means that the operators $T_a$ and $T_b$ defined as $T_a x := a \circ x$ and $T_b x := b \circ x$ commute and therefore $U_a = 2T_a^2 - T_a$ and $U_b = 2T_b^2 - T_b$ commute such that we have the equivalence of (vii) and (viii).

From (xi) we get (viii) by Lemma 5.2 and (x) by considering the subalgebra $\mathbb{R} d_1 + \mathbb{R} d_2 + \mathbb{R} d_3$; this subalgebra is associative and contains $e = d_1 + d_2$ as well as $f = d_2 + d_3$. Now suppose (x). Then $e$, $f$ and $\mathbb{1}$ generate an associative subalgebra such that $(a \circ b)^2 = a^2 \circ b^2 = a \circ b$ for $a,b \in \{e,e',f,f'\}$, and we have (ix). From (ix) we get (xi) by defining $d_1 := e \circ f'$, $d_2 := e \circ f$ and $d_3 := e' \circ f$. Then $d_1$ and $d_2$ are orthogonal since $d_1 \le d_1 + e' \circ f' = f'$ and $d_2 \le d_2 + d_3 = f$; $d_2$ and $d_3$ are orthogonal since $d_2 \le d_1 + d_2 = e$ and $d_3 \le d_3 + e' \circ f' = e'$; $d_1$ and $d_3$ are orthogonal since $d_1 \le d_1 + d_2 = e$ and $d_3 \le e'$.             q.e.d.

The identity $\mu(f) = \mu(e)\mu(f|e) + \mu(e')\mu(f|e')$ is a well-known rule for classical conditional probabilities. However, it is not anymore universally valid in a nonclassical framework like quantum mechanics. Its validity for all states $\mu$ becomes a first weak and asymmetrical notion of compatibility for a pair of events $e$ and $f$. This is condition (i) and is equivalent to the algebraic condition (ii).

Its validity also for exchanged roles of $e$ and $f$ becomes a stronger and symmetrical notion of compatibility. This is condition (iii) of the above proposition. It is equivalent to each one of the conditions (iv), (v) and (vi). The latter one represents another rule for classical conditional probabilities which is not anymore universally valid in quantum mechanics: $\mu(a)\mu(b|a) = \mu(b)\mu(a|b)$ for $a,b \in \{e,e',f,f'\}$ and $\mu \in S$.

A still stronger form of compatibility is described by each one of the two equivalent conditions (vii) and (viii). While (viii) represents a purely algebraic condition, (vii) has an interesting interpretation in quantum measurement; it means that the order of two successive measurements in a series of measurements does not matter when one of the two successive measurement tests $e$ versus $e'$ and the other one $f$ versus $f'$. For a deeper look at this, iterated conditional probabilities and their connection to quantum measurement must be considered (see [5]). If each pair of elements in $E$ satisfies (vii), then $A$ is associative by (viii) which implies (x) such that all conditions are satisfied in this case, resulting in a classical situation.

The strongest level of compatibility for two events $e$ and $f$ is represented by each one of the three equivalent conditions (ix), (x) and (xi). The latter one means that $e$ and $f$ lie in a Boolean subalgebra of $E$.

If $A$ is the self-adjoint part of a W*-algebra, the weakest one among all the conditions which is (i) means $f = efe + e'fe'$ and implies $ef = efe = fe$ such that $e$ and $f$ commute and (ix) holds. Therefore, in this case, all the above conditions become equivalent and there is only one single level of compatibility coinciding with the usual concept of commuting operators in quantum mechanics. Likewise (i) implies (ix) and all conditions become equivalent if $A$ is a JBW algebra [2]. In the more general framework of the assumptions 3.1, however, there appear to be four different levels of compatibility. For each level there is a set of equivalent conditions describing it; these sets are: (i)-(ii), (iii)-(vi), (vii)-(viii), (ix)-(xi). The verification that these levels really differ still requires an example of an algebra that satisfies the assumptions 3.1, but is not a JB algebra.

**Corollary 5.4:** *If an associative subalgebra M of A is generated by its idempotent elements, then $y^2 \geq 0$ for $y \in M$.*

*Proof.* Any $e, f \in E \cap M$ satisfy (x) in Proposition 5.3 and thus (ix) and (xi). Therefore each element $y \in M$ has the shape $y = \Sigma t_k e_k$ with real numbers $t_k$ and orthogonal elements $e_1, ..., e_n \in E \cap M$. Then $y^2 = \Sigma t_k^2 e_k \geq 0$. 　　　　q.e.d.

## 6. Matrix algebras

Let $R$ be a real *-algebra with unit 1 and define $R_{sa} := \{\alpha \in R : \alpha = \alpha^*\}$. Note that the product in $R$ is neither assumed to be commutative nor associative nor alternative. Let $H_n(R)$ denote the space of Hermitian, or self-adjoint, $n \times n$ matrices with coefficients in $R$. On $H_n(R)$ consider the product defined by $a \circ b := (ab + ba)/2$. This type of Hermitian matrix algebras is studied in the present section because they are a natural candidate for a structure satisfying the assumptions 3.1.

A case of particular interest is when the *-algebra $R$ is the octonions, since then $H_n(R)$ is a formally real Jordan algebra for $n \leq 3$, but not for $n \geq 4$ [2]. Therefore one might hope to find with $n \geq 4$ the desired example that satisfies the assumptions 3.1, but is not a JB algebra. This will turn out to be false. The results in this section will even be a lot more general. We shall see that any matrix algebra satisfying the assumptions 3.1 must be a JB algebra.

**Lemma 6.1:** (i) *If $R_{sa} = \mathbb{R}$ (note that $\mathbb{R} 1$ is identified with $\mathbb{R}$ here), then $\alpha \alpha^* = \alpha^* \alpha$ for $\alpha \in R$. If furthermore $\alpha^* \alpha \neq 0$ holds for each $\alpha \neq 0$, then we have $\alpha^* \alpha \geq 0$ for $\alpha \in R$.*
(ii) *If $R_{sa} = \mathbb{R}$ and if $\alpha^* \alpha \neq 0$ holds for each nonzero element $\alpha$ in $R$, then $H_2(R)$ is a norm-dense subalgebra of a spin factor.*

*Proof.* (i) Assume $R_{sa} = \mathbb{R}$ and $\alpha \in R$. Then $t := \alpha + \alpha^* \in \mathbb{R}$ such that $\alpha$ and $\alpha^* = t - \alpha$ commute. Now assume $\alpha^* \alpha \neq 0$ for all $\alpha \neq 0$ and $\beta^* \beta < 0$ for some $\beta \in R$. Consider the real polynomial function $h(s) := (s + (1-s)\beta)^*(s + (1-s)\beta) = s^2 + s(1-s)(\beta + \beta^*) + (1-s)^2 \beta^* \beta$. Since $h(0) = \beta^* \beta < 0$ and $h(1) = 1$, there exists some $s_o$ with $0 < s_o < 1$ and $h(s_o) = 0$ such that $\alpha^* \alpha = 0$ for $\alpha = s_o + (1-s_o)\beta$. Therefore $s_o + (1-s_o)\beta = 0$, thus $\beta \in \mathbb{R}$ and $\beta^* \beta = \beta^2 \geq 0$.

(ii) Assume $n=2$, $R_{sa} = \mathbb{R}$ and $\alpha^* \alpha \neq 0$ for $0 \neq \alpha \in R$. Denote by $a_{ij}$ the matrix whose entry in the $i$-th row and $j$-th column is 1 and whose all other entries are zero. Define $V := \{\alpha(a_{11} - a_{22}) + \beta a_{12} + \beta^* a_{21} : \alpha \in \mathbb{R}, \beta \in R\} \subseteq H_2(R)$. Then $H_2(R)$ is the direct sum of $V$ and $\mathbb{R}$. Moreover, for $x = \alpha_1(a_{11} - a_{22}) + \beta_1 a_{12} + \beta_1^* a_{21} \in V$ and $y = \alpha_2(a_{11} - a_{22}) + \beta_2 a_{12} + \beta_2^* a_{21} \in V$ with $\alpha_1, \alpha_2 \in \mathbb{R}$, $\beta_1, \beta_2 \in R$, we get from (i) that $x \circ y = \alpha_1 \alpha_2 + (\beta_1 + \beta_2)^*(\beta_1 + \beta_2) - \beta_1^* \beta_1 - \beta_2^* \beta_2$ and thus $x \circ y \in \mathbb{R}$, $x \circ x \geq 0$ and $x \circ x \neq 0$ for $x \neq 0$. Therefore, $x \circ y$ becomes an inner product on $V$. If the real dimension of $R$ is finite, so is the one of $V$ and $H_2(R)$ is a finite-dimensional spin factor. If the dimension of $R$ is not finite, neither $R$ nor $V$ need be complete. In this case, let $W$ be the completion of the pre-Hilbert space $V$ and consider the spin factor $W \oplus \mathbb{R} \mathbb{1}$. Then $H_2(R)$ becomes a norm-dense subalgebra of this spin factor. 　　　　q.e.d.

**Lemma 6.2:** *Suppose that $A=H_n(R)$ and $E=\{e \in A: e^2=e\}$ satisfy the assumptions* 3.1.
(i) *If $\alpha=\alpha^2$ holds only for $\alpha=0$ and $\alpha=1$ in $R_{sa}$ (i.e., $R_{sa}$ does not contain nontrivial idempotents), then $R_{sa}=\mathbb{R}$.*
(ii) *If $n\geq 2$, then $R$ must not contain any element $\alpha$ with $\alpha^*\alpha=\alpha\alpha^* \in \mathbb{R}$ and $\alpha^*\alpha=\alpha\alpha^*<0$. Particularly, $R_{sa}$ must not contain any element $\alpha$ with $\alpha^2 \in \mathbb{R}$ and $\alpha^2<0$. Moreover, $\alpha^*\alpha \neq 0$ or $\alpha\alpha^* \neq 0$ for $\alpha \neq 0$. If $R_{sa}=\mathbb{R}$, we have $\alpha^*\alpha=\alpha\alpha^*>0$ for $\alpha \neq 0$.*
(iii) *If $n\geq 3$ and $R_{sa}=\mathbb{R}$, then $R$ is alternative and does not contain any zero divisors.*
(iv) *If $n\geq 4$, then $R$ is associative (and $A=H_n(R)$ becomes a special Jordan algebra* [2]*).*

*Proof.* Denote by $a_{ij}$ the matrix whose entry in the *i*-th row and *j*-th column is 1 and whose all other entries are zero.

(i) Suppose that $\alpha=\alpha^2$ holds only for $\alpha=0$ and $\alpha=1$ in $R_{sa}$. Consider $e:=a_{11} \in E$. If $d \in E$ with $d \leq e$, then $d=U_e d=\{e,d,e\}=\alpha e$, where $\alpha \in R_{sa}$ is the first entry in the first row of the matrix $d$. Thus $\alpha e=d=d^2=\alpha^2 e$ and $\alpha=\alpha^2$ such that either $\alpha=0$ and $d=0$ or $\alpha=1$ and $d=e$. Therefore $U_e A = \overline{\text{lin}}\{d \in E: d \leq e\} = \mathbb{R} e$. For any $\alpha \in R_{sa}$ now consider the matrix $\alpha e$ in $A$ and conclude $\alpha \in \mathbb{R}$ from $\alpha e=\{e,\alpha e,e\}=U_e(\alpha e) \in \mathbb{R} e$.

(ii) Suppose $n \geq 2$ and $\alpha^*\alpha=\alpha\alpha^*<0$ for an element $\alpha \in R$. Without loss of generality assume that $\alpha^*\alpha = \alpha\alpha^* = -1$ (if this is not the case, replace $\alpha$ by $\alpha/(-\alpha^*\alpha)^{1/2}$). Then consider the two matrices $e:=a_{11}$ and $f:=\frac{1}{2}(1-5^{1/2})a_{11}+\frac{1}{2}(1+5^{1/2})a_{22}+\alpha a_{12}+\alpha^* a_{21}$. Both matrices are idempotent and thus lie in $E$, but $U_e f=\{e,f,e\}=\frac{1}{2}(1-5^{1/2})e$ is not positive since $\frac{1}{2}(1-5^{1/2})<0$.

Now assume $n \geq 2$ and $\alpha^*\alpha=\alpha\alpha^*=0$ for an element $\alpha \in R$. Consider $x:=\alpha a_{12}+\alpha^* a_{21}$. Then $x^2=0$ and $e \circ x = x/2$. Therefore $(e+sx)^2=e+sx$ such that $e+sx \in E$ and $e+sx \leq \mathbb{I}$ for all $s \in \mathbb{R}$. Thus $sx \leq \mathbb{I}$ for all $s \in \mathbb{R}$ and $x=0$ such that $\alpha=0$. The remaining part of (ii) follows from Lemma 6.1.

(iii) Suppose $n \geq 3$ and $R_{sa}=\mathbb{R}$. For any two elements $\alpha,\beta \in R$ with $\alpha^*\alpha=1$ consider the following four matrices: $e:=(a_{11}+a_{22}+\alpha a_{12}+\alpha^* a_{21})/2$, $x:=\beta a_{23}+\beta^* a_{32}$, $d:=a_{11}+a_{22}$ and $f:=a_{33}$. Then $e,d,f \in E$, $e \leq d$, and $d$ and $f$ are orthogonal. Therefore $f$ and each $g \in E$ with $g \leq e$ are orthogonal such that $f \circ g = 0$ by Lemma 3.3. From $U_e x \in \overline{\text{lin}}\{g \in E: g \leq e\}$ we get that $f \circ U_e x=0$. Multiplying out the matrices gives first $8U_e x=[\alpha^*(\alpha\beta)-\beta]a_{23}+[\alpha^*(\alpha\beta)-\beta]^*a_{32}$ and finally $16 f \circ U_e x = [\alpha^*(\alpha\beta)-\beta]a_{23}+[\alpha^*(\alpha\beta)-\beta]^*a_{32}$ such that $\alpha^*(\alpha\beta)=\beta$.

For $\alpha \neq 0$ we can apply this to $\alpha/(\alpha^*\alpha)^{1/2}$ and get $\alpha^*(\alpha\beta)=(\alpha^*\alpha)\beta$. This identity also holds for $\alpha=0$. Since $\alpha+\alpha^* \in R_{sa}=\mathbb{R}$, we have $(\alpha+\alpha^*)(\alpha\beta) = ((\alpha+\alpha^*)\alpha)\beta$ and thus $\alpha(\alpha\beta)+\alpha^*(\alpha\beta) = \alpha^2\beta+(\alpha^*\alpha)\beta$. Combining this with the other identity yields $\alpha(\alpha\beta)=\alpha^2\beta$. This is the left alternative law. Taking adjoints, we obtain the right alternative law.

Since $R$ is alternative, each pair of elements in $R$ generates an associative subalgebra. Therefore, for any two elements $\alpha,\beta \in R$, all four elements $\alpha,\alpha^*,\beta,\beta^*$ lie in the associative subalgebra generated by $\alpha-\alpha^*$ and $\beta-\beta^*$ because $\alpha+\alpha^*$ and $\beta+\beta^*$ are real numbers. Thus $(\alpha\beta)(\alpha\beta)^* = (\alpha\beta)(\beta^*\alpha^*) = \alpha(\beta\beta^*)\alpha^* = (\alpha\alpha^*)(\beta\beta^*)$. If now $\alpha\beta=0$, $\alpha=0$ or $\beta=0$ must hold.

(iv) Suppose $n \geq 4$. For $\alpha,\beta,\gamma \in R$ consider the following four matrices in $H_n(R)$: $x:=\alpha a_{12}+\alpha^* a_{21}$, $y:=\beta a_{23}+\beta^* a_{32}$, $z:=\gamma a_{34}+\gamma^* a_{43}$, and $e:=a_{11}+a_{22}$. Then $e \in E$, $x \in \{e,A,e\}=U_e A$, and $z \in \{e',A,e'\}=U_{e'}A$. Lemma 5.2 implies $x \circ (z \circ y)=z \circ (x \circ y)$. Multiplying out the matrices, we get $\alpha(\beta\gamma)=(\alpha\beta)\gamma$.                                                                q.e.d.

When Lemma 5.2 is available, the proof of part (iv) of the above lemma becomes an exact copy of the one of the same result for Jordan algebras [2], but the proof of (iii) is different.

**Theorem 6.3:** *Suppose that $A=H_n(R)$ and $E=\{e \in A: e^2=e\}$ satisfy the assumptions* 3.1 *with a real \*-algebra $R$ with unit $1$ and $R_{sa}=\mathbb{R}$. Then $A$ is a formally real Jordan algebra. If $n=2$, $A$ is a spin factor. If $n=3$, $R$ is the real numbers, the complex numbers, the quaternions or the octonions. If $n \geq 4$, $R$ is the real numbers, the complex numbers or the quaternions.*

*Proof.* If $n=1$, $A=H_1(R)=\mathbb{R}$. If $n=2$, $A=H_2(R)$ is a spin factor by Lemma 6.1 (ii); the completeness follows from the assumptions 3.1 since dual spaces are Banach spaces. If $n=3$, $R$ is an alternative real division algebra by Lemma 6.2 (iii), and there are no other such algebras than the real numbers, the complex numbers, the quaternions or the octonions [1]. If $n \geq 4$, $R$ is an associative real division algebra by Lemma 6.2 (iii) and (iv), and there are no other such algebras than the real numbers, the complex numbers and the quaternions [1]. In all cases, $A=H_n(R)$ is a formally real Jordan algebra [2], if the involution * on $R$ coincides with the usual conjugation on these division algebras. This conjugation is characterized by linearity over $\mathbb{R}$ and the requirements $1^*=1$ and $j^*=-j$ whenever $j^2=-1$. The identity $1^*=1$ follows from $R_{sa}=\mathbb{R}$. Now suppose $j^2=-1$ and define $t:=j^*j \in \mathbb{R}$. Using the right alternative lay, we get $-tj = -(j^*j)j = -j^*j^2 = j^*$ and taking adjoints $j = -tj^* = t^2 j$, so that $t=\pm 1$ and $j^*=\pm j$. However, $j^*=j$ is impossible since then $j \in \mathbb{R}$ such that $j^2 \neq -1$.     q.e.d.

We now look at some further consequences of the above lemmas. Besides the real numbers, complex numbers, quaternions and octonions, there are some more *-algebras, the potential relevance of which for modern physics is sometimes discussed. A natural question is therefore whether $H_n(R)$ satisfies the assumptions 3.1 when $R$ is one of these *-algebras. The bioctonions, quateroctonions, octooctonions, which are linked with the exceptional Lie groups by the so-called magic square [1], do not satisfy the condition $R_{sa}=\mathbb{R}$ and are thus not covered by Theorem 6.3. However, they are ruled out by Lemma 6.2 (ii). A further type of *-algebras arises when the Cayley-Dickson construction [1] is continued beyond the octonions. This case is covered and ruled out by Theorem 6.3 for $n \geq 3$ and results in spin factors for $n=2$ by Lemma 6.1 (ii). The split-complex numbers, split-quaternions, split-octonions are excluded by Theorem 6.3 for $n \geq 3$ and by Lemma 6.2 (ii) for all cases $n \geq 2$.

## 7. Conclusions

The structure considered here is motivated by the results in [6] where an abstract setting for conditional probabilities and the Lüders-von Neumann quantum measurement was introduced. The assumptions 3.1 do not represent the most general case of this setting, but a more specialized case where the conditional probabilities can be expressed by a nonassociative product as it is possible in the Jordan operator algebras. This specialized case is ideally suited for studying the question whether matrix algebras can provide examples of that setting.

We have seen that an algebra $A$ satisfying the assumptions 3.1 need not necessarily be a Jordan algebra, but still features some important properties of an operator algebra. This includes existence and uniqueness of the conditional probabilities as well as the uniqueness of the spectral resolution. However, the spectral resolution does not exist for all elements of the algebra, but only for those which generate an associative subalgebra with positive squares.

From a mathematical point of view, such an algebra $A$ might be an interesting generalization of an operator algebra. From a physical point of view, it might provide a framework for the study of quantum measurement in a non-Hilbert-space environment to reveal those characteristics of the quantum measurement process which depend on the Hilbert space formalism and those which do not.

Supposing that the elements of $E=\{e \in A: e^2=e\}$ represent quantum events, three different cases can be distinguished for two events $e$ and $f$: (1) $e$ and $f$ belong to an associative subalgebra with positive squares, (2) $e$ and $f$ belong to a power-associative subalgebra with positive squares, and (3) none of these first two cases holds. Then the first case represents the well-known case of compatible or commuting events resulting in a classical situation. The second case represents the standard situation in quantum mechanics. The third case, however, is new and might exhibit properties unknown from classical theories as well as quantum mechanics. The four different levels of compatibility discussed in section 4 might be a first hint.

However, the existence of this third case is unproved as long as an example of an algebra $A$ that satisfies the assumptions 3.1, but is not a JB algebra has not been found. The present paper contributes to the search for such an example by analyzing some natural first candidates (a certain type of "small" algebras, matrix algebras), but they all turn out to become JB algebras if they satisfy the assumptions 3.1. Moreover, some major differences between potential examples and JB algebras have been identified; either power-associativity or the positivity of the squares or both will not hold in the desired example.

The possibility that no such example exists cannot be ruled out. However, it would mean that the assumptions 3.1 provide an unexpected new characterization of JBW algebras, which would have a significant impact on quantum axiomatics since some customary axioms (e.g., power-associativity or the sum postulate for observables [3,5,6]) might become redundant then.